\documentclass{article}

\usepackage{enumerate}
\usepackage{graphicx}
\usepackage{amsmath}
\usepackage{amssymb}
\usepackage{amsthm}
\usepackage{multicol}
\usepackage{mdframed}
\usepackage{framed}
\usepackage{url}
\usepackage{hyperref}

\newenvironment{tops}{\framed\noindent\textbf{\textsf{Tamper Opportunities
      Problem (TOP):\\}}}{\endframed}
\newenvironment{tsps}{\framed\noindent\textbf{\textsf{Tamper Strategies
      Problem (TSP):\\}}}{\endframed}
\newenvironment{epps}{\framed\noindent\textbf{\textsf{Evidence-Protecting Program
      Problem (EPPP):\\}}}{\endframed}
\newenvironment{labeledbox}[1]{\framed\noindent\textbf{\textsf{#1:}}}{\endframed}
\usepackage[dvipsnames]{xcolor}
\graphicspath{ {./} }

\newcommand{\ts}[1]{\ensuremath{\fnf{ts}(#1)}}

\newcommand{\etp}{\ensuremath{\tau}}

\newcommand{\opps}{\ensuremath{\cnc{Opps}}}
\newcommand{\evisc}{\cnf{epp}}
\newcommand{\mypar}[1]{\vspace{1ex}\noindent \textbf{#1}~}

\usepackage{url}
\usepackage{listings}
\lstnewenvironment{code}%
    {\lstset{}%
      \csname lst@SetFirstLabel\endcsname}
    {\csname lst@SaveFirstLabel\endcsname}
\newtheorem{thm}{Theorem}
\newtheorem{lem}{Lemma}
\newtheorem{cor}{Corollary}
\newtheorem{defn}{Definition}
\newtheorem{xmpl}{Example}

\newcommand{\cnf}[1]{\ensuremath{\operatorname{\mathsf{#1}}}}
\newcommand{\cnc}[1]{\ensuremath{\mathsf{#1}}}
\newcommand{\fnf}[1]{\ensuremath{\operatorname{\mathit{#1}}}}

\newcommand{\seq}[1]{\ensuremath{\langle#1\rangle}}
\newcommand{\app}{\mathbin{\ast}}
\newcommand{\cons}{\mathbin{::}}
\newcommand{\length}[1]{\ensuremath{|#1|}}

\newcommand{\append}{\mathbin{\S}}

\newcommand{\nil}{\mathord{-}}
\newcommand{\cpy}{\mathord{+}}
\newcommand{\nullify}{\ensuremath{\{\}}}
\newcommand{\copyit}{\ensuremath{{}_{\displaystyle -}}}
\newcommand{\sign}{\ensuremath{\mathord !}}
\newcommand{\hash}{\ensuremath{\mathord\#}}
\newcommand{\linseq}{\mathord{<}}
\newcommand{\linpar}{\mathord{\sim}}
\newcommand{\linseqe}{\to}
\newcommand{\braseqe}[2]{\mathbin{{#1}\linseq{#2}}}
\newcommand{\brapare}[2]{\mathbin{{#1}\linpar{#2}}}
\newcommand{\at}[2]{\mathop{@_{#1}}{[#2]}}

\newcommand{\mt}{\xi}
\newcommand{\signp}[2]{\mathop{!}\{#1\}_{#2}}
\newcommand{\hashp}[2]{\mathop{\#}\{#1\}_{#2}}
\newcommand{\sseq}{\mathbin{;\!;{}}}
\newcommand{\spar}{\parallel}

\newcommand{\eval}[4]{\mathcal{E}({#1},{#2},{#3},{#4})}
\newcommand{\evalC}[1]{\mathcal{C}({#1})}
\newcommand{\evalF}[2]{\mathcal{F}({#1},{#2})}

\newcommand{\dataC}[1]{\bar{\mathcal{C}}({#1})}
\newcommand{\dataE}[4]{\bar{\mathcal{D}}({#1},{#2},{#3},{#4})}

\newcommand{\dataS}[1]{\bar{\mathcal{S}}({#1})}
\newcommand{\bfr}[1]{\mathbin{\rhd{#1}}}
\newcommand{\bfrnil}{\bfr{\nil}}
\newcommand{\bfrcpy}{\bfr{\cpy}}

\newcommand{\projp}{\varrho}
\newcommand{\dprojp}{\projp^{\prime}}
\newcommand{\proje}{\varepsilon}

\newcommand{\tmprs}[2]{\overline{\Pi}({#1},{#2})}

\title{Evidence Tampering and Chain of Custody in Layered Attestations}
\author{Ian D. Kretz \and Clare C. Parran \and John D. Ramsdell \and
  Paul D. Rowe\thanks{Affiliated with MITRE when this work was done.} \\
  \texttt{$\{$ikretz, ccparran, ramsdell$\}$@mitre.org} \\
  \texttt{paul.rowe@twosixtech.com}}
\date{}

\begin{document}
\maketitle

\newcommand\blfootnote[1]{%
  \begingroup
  \renewcommand\thefootnote{}\footnote{#1}%
  \addtocounter{footnote}{-1}%
  \endgroup
}

\blfootnote{Approved for Public Release; Distribution Unlimited.
  Public Release Case Number 23-2478.  This technical data was
  developed using contract funds under Basic Contract
  No.~W56KGU-18-D-0004. The view, opinions, and/or findings contained
  in this report are those of The MITRE Corporation and should not be
  construed as an official Government position, policy, or decision,
  unless designated by other documentation.\\ \copyright~2023 The
  MITRE Corporation. This paper is licensed under the terms of the
  Creative Commons Attribution 4.0 International License
  (\url{http://creativecommons.org/licenses/by/4.0/}).}

\begin{abstract}
  In distributed systems, trust decisions are made on the basis of
  integrity evidence generated via remote attestation. Examples of the
  kinds of evidence that might be collected are boot time image hash
  values; fingerprints of initialization files for userspace
  applications; and a comprehensive measurement of a running kernel.
  In layered attestations, evidence is typically composed of
  measurements of key subcomponents taken from different trust
  boundaries within a target system. Discrete measurement evidence is
  bundled together for appraisal by the components that collectively
  perform the attestation.

  In this paper, we initiate the study of evidence chain of custody
  for remote attestation.  Using the Copland attestation specification
  language, we formally define the conditions under which a runtime
  adversary active on the target system can tamper with measurement
  evidence.  We present algorithms for identifying all such tampering
  opportunities for given evidence as well as tampering ``strategies''
  by which an adversary can modify incriminating evidence without
  being detected.  We then define a procedure for transforming a
  Copland-specified attestation into a maximally tamper-resistant
  version of itself.  Our efforts are intended to help attestation
  protocol designers ensure their protocols reduce evidence tampering
  opportunities to the smallest, most trustworthy set of components
  possible.
\end{abstract}


\section{Introduction}\label{sec:intro}

Chain of custody in criminal investigations is crucial because it
serves to convince a jury that the evidence presented can be traced
directly back to its collection at a crime scene. The chain of custody
limits and documents who could be responsible for any discrepancy
between the evidence collected and the evidence presented.

Remote attestation shares similarities with a criminal
investigation. In a remote attestation, evidence of a target's
integrity is collected by measuring various subcomponents of the
target.  The evidence can be collected in a variety of ways.  Some
examples are:

\begin{itemize}
\item the hashes of images used to boot an operating system can be
  digested into a report;
\item a fingerprint of the initialization files of a critical
  userspace application can make a measurement; and
\item an extensive analysis of a running kernel can be reduced to an
  informative measurement using a system such as
  LKIM~\cite{Loscocco:07:Linux-kernel-in}.
\end{itemize}

This evidence is bundled together and presented to an appraiser who
evaluates the evidence to infer whether the target has sufficient
integrity. Thus, chain of custody for measurement evidence is also
crucial for remote attestation. It should convince the appraiser that
the evidence it receives can be traced directly back to the act of
measurement itself. Any discrepancies can be attributed to whoever had
access to the evidence along the chain of custody. The consequence of
blindly trusting a component without sufficient integrity guarantees
can have limitless repercussions since it would give an adversary a
critical foothold on the system.

In this paper, we initiate the study of chain of custody for remote
attestations. We focus on the following questions which are important
for an appraiser to ask.
\begin{quote}
  Which components had an opportunity to tamper with a piece of
  evidence and remove signs of corruption?
\end{quote}
\begin{quote}
  Could an adversary have combined these opportunities to eliminate
  all copies of a piece of incriminating evidence prior to appraisal?
\end{quote}
\begin{quote}
  Can tampering opportunities be eliminated by enhancing an
  attestation's design?
\end{quote}

Our primary goal is to help designers of remote attestation protocols
ensure their protocols limit these evidence tampering opportunities
as much as possible.

An important aspect of our approach is
Copland~\cite{RamsdellEtAl2019}, a language for specifying layered
attestation protocols. Layered attestations take advantage of the
hierarchical nature of many system architectures where measurements
may be taken from a variety of \emph{places} that may have
different inherent levels of protection and different dependencies.
Copland is equipped with formal semantics that allows for rigorous
formal analyses of trust properties~\cite{RoweRK2021} while remaining
connected to concrete implementations~\cite{PetzA2021, MaatArxiv2017,
  Maat2018}.

Copland allows us to introduce and formalize three problems
corresponding to these questions.

\begin{description}
\item[Tamper Opportunities Problem (TOP):] Given a Copland phrase and a
  measurement event $v$ in its semantics, find all tampering
  opportunities for the measurement evidence produced at $v$.
\end{description}

\begin{description}
\item[Tamper Strategies Problem (TSP):] Given a Copland phrase and a
  measurement event $v$ in its semantics, find all adversarial
  ``strategies'' for tampering that consistently modify every copy
  of~$v$ that the appraiser sees.
\end{description}

\begin{description}
\item[Evidence-Protecting Program Problem (EPPP):] Given a Copland
  phrase, produce another that preserves the given phrase's semantics
  while maximally constraining the set of tampering opportunities
  available to an adversary.
\end{description}

The major contributions of this paper are:
\begin{enumerate}
\item Initiating the study of chain of custody for layered attestations
\item Formalizing the TOP and TSP and introducing algorithms for
  computing solutions to them
\item Formalizing the EPPP and introducing an idempotent function that
  maps each Copland phrase to a maximally tamper-proof iteration of
  itself
\end{enumerate}

All definitions and results of the paper have been formalized in the
Coq proof assistant~\cite{Coq18}. The proof scripts are available from
the public repository~\url{https://github.com/mitre/copland}.

The remainder of the paper is structured as
follows. Section~\ref{sec:copland} introduces the syntax and semantics
of the Copland language.  Section~\ref{sec:motivation} motivates the
study of chain of custody for layered attestations.
Section~\ref{sec:tampering} formalizes the TOP and TSP and defines
high-level algorithms for solving them.  In Section~\ref{sec:results}
we introduce the notion of a \emph{protected} data flow graph and
prove that such graphs minimize tamper opportunities available to an
adversary. The Copland Evidence Protection Program is defined in
Section~\ref{sec:adding signatures} and proven to solve the EPPP for
any input phrase.  We discuss how our work relates to other efforts
from the literature in Section~\ref{sec:related} before concluding.
Supporting definitions for the Copland Data Flow Semantics can be
found in Appendix A.


\section{Copland Syntax and Semantics}\label{sec:copland}

Copland is a domain-specific language for specifying layered
attestations. The semantics of Copland is fundamental to our
understanding of what it means for an adversary to be able to tamper
with evidence during an attestation. This section presents the syntax
and semantics of Copland together with a small set of notational
conventions. Ambitious readers can proceed straight to the motivating
examples of Section~\ref{sec:motivation}, returning here as needed for
technical reference.

\begin{figure}
  \[\begin{array}{rcl@{\quad}l}
      C&\gets&\ast P : T&\mbox{Start $T$ at $P$}\\
      T&\gets&S~P~S&\mbox{Measurement (Probe Place Target)}\\
       &\mid&@_P~[~T~]&\mbox{At place}\\
       &\mid&$\copyit$&\mbox{Copy}\\
       &\mid&$\sign$&\mbox{Sign}\\
       &\mid&$\hash$&\mbox{Hash}\\
       &\mid&(T~\linseqe~T)&\mbox{Linear sequence}\\
       &\mid&(T~\braseqe{D}{D}~T)&\mbox{Sequential branching}\\
       &\mid&(T~\brapare{D}{D}~T)&\mbox{Parallel branching}\\
      D&\gets&\nil\mid\cpy&\mbox{Splitting specification}
    \end{array}
  \]

  \caption{Copland Syntax}\label{fig:copland syntax}
\end{figure}

The syntax of Copland is presented in Figure~\ref{fig:copland
  syntax}. The syntactic category $P$ specifies places at which
Copland actions take place, the category $S$ specifies symbols that
identify measurements or targets of measurements, $C$ is the start
symbol and the $\linseqe$ operator is right-associative.

The data flow semantics of a Copland phrase is given as a \emph{data
flow graph}, a labeled, directed acyclic graph with distinguished
input and output nodes.  The nodes represent events such as measuring
a component, signing evidence, or making a remote request; the edges
encode data and control flow.

\begin{defn}[Data Flow Graph]\label{def:data flow graph}
  A \emph{data flow graph} $(V,v^i,v^o,\to,\ell)$ consists~of
  \begin{multicols}{2}
  \begin{enumerate}
  \item set $V$ of events,
  \item $v^i\in V$, the input event,
  \item $v^o\in V$, the output event,
  \item relation $\to\subseteq
    V\times V$,
    flow edges,
  \item function $\ell:V\to L$, a map from events to labels.
  \end{enumerate}
  \end{multicols}
  The graph $(V,\to)$ must be acyclic, and there must be no flow edge
  to $v^i$ or from $v^o$.

  For $D=(V,v^i,v^o,\to,\ell)$ we define $D^V=V$,
  $\bot(D)=v^i$, and $\top(D)=v^o$.
\end{defn}

Isolating the data flow portion of the semantics allows us to reason
about how evidence propagates from component to component as an
attestation executes.  Such analyses form the basis of our ability to
identify tampering opportunities.

Figure~\ref{fig:data flow semantics} in the appendix defines the data
flow semantics inductively on the structure of Copland phrases.  In
this section, we seek to illustrate how the semantics uses this
structure to build up a data flow graph.  We rely on diagrams of the form

\begin{center}
  \includegraphics{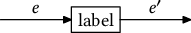}
\end{center}

\noindent to describe how executing a Copland phrase $c$ at place $p$
transforms and transmits evidence.  Boxes represent labeled events and
arrows depict the flow of evidence and the precedence ordering of
events.  The syntax of evidence is given in Figure~\ref{fig:evidence
  syntax}.  The syntax of event labels is given in Figure~\ref{fig:label
  syntax}.  Labels identify what kind of action (e.g., $\cnc{msp}$ for
measurement, $\cnc{sig}$ for signature, $\cnc{req}$ for request) an
event represents, the place where the action occurs, the evidence
produced at the event, and other action-specific data.  These diagrams
and the data flow semantics also refer to the Copland evidence
semantics $\mathcal{E}$ defined in Figure~\ref{fig:evidence
  semantics}.  This semantics computes the form of the evidence that
is produced by executing a Copland phrase at a place.

\begin{figure}

  \[\begin{array}{rcl@{\quad}l}
  E&\gets&\mt&\mbox{Empty evidence}\\
  &\mid&\cnf{m}(\cnf{ms}(S,P,S),P,\Phi,E)&\mbox{Measurement}\\
  &\mid&\signp{E}{P}&\mbox{Signature}\\
  &\mid&\hashp{E}{P}&\mbox{Hash}\\
  &\mid&(E\sseq E)&\mbox{Sequential composition}\\
  &\mid&(E\spar E)&\mbox{Parallel composition}\\
  \Phi&\gets&\mbox{a sequence of positive integers}
  \end{array}\]

  \caption{Evidence Syntax}\label{fig:evidence syntax}

\end{figure}

\begin{figure}
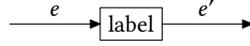

  \[\begin{array}{rcl}
      L&\gets&P:\cnf{msp}(E)
               \mid P:\cnc{cpy}(E)\mid P:\cnc{sig}(E)\\
       &\mid&P:\cnc{hsh}(E)\mid P:\cnc{req}(P, E)\mid P:\cnf{rpy}(P, E)\\
       &\mid&P:\cnf{split}(D,D,E)\mid P:\cnf{join}(O,E)\\
      O&\gets&\linseq\mid\linpar
  \end{array}\]

  \caption{Label Syntax}
  \label{fig:label syntax}
\end{figure}

The most basic Copland phrase is a measurement $m~q~t$: symbols $m$
and $t$ name the probe and the target of measurement, respectively,
and $q$ is the place where $t$ resides.  When at place $p$, $m~q~t$
means that $p$ should receive some evidence $e$, perform $m$ targeting
$t$ at $q$, and then emit the resulting evidence.  The structure of
evidence produced by the measurement $m~q~t$ is $\cnf{msp}(\cnf{m}(p,
m, q, t, v_{\phi}, e))$.  When looking at the abstract syntax of a
Copland phrase, $\phi$ provides a path from the root to the position
of $m~q~t$ in the phrase, $v_{\phi}$ is a variable that holds the raw
evidence value produced by the measurement, and $e$ is the evidence
term passed as input to the measurement event.  The path is used to
produce distinguished evidence when two occurrences of a measurement
request in a phrase have the same measurement, target and place.

\begin{center}
  \includegraphics{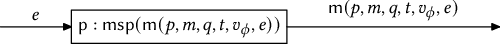}
\end{center}

The semantics of the phrases copy $\copyit$, sign $\sign$, and
hash~$\hash$ have the same form as a measurement.  When executing at
place $p$, their corresponding event labels are $p:\cnc{cpy}(e)$,
$p:\cnc{sig}(e)$, and $p:\cnc{hsh}(e)$. For input evidence $e$, the
evidence sent along the outgoing arrow is $e$, $\signp{e}{p}$, and
$\hashp{e}{p}$, respectively.

The semantics of the remainder of the Copland syntax is defined
inductively.  Let $p:\{c\}$ be the events and their labels and
orderings associated with executing phrase $c$ at $p$, and let
$\eval{c}{p}{\phi}{e}$ be the evidence that results from the execution
when $e$ is provided as input evidence and when $c$ is located at
position $\phi$ inside a Copland phrase.  Measurements can be combined
in a pipeline fashion using the $\linseqe$ operator.  Thus, when at
$p$, $c_1\linseqe c_2$ means

\begin{center}
  \includegraphics{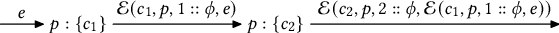}
\end{center}

A measurement can be taken at a remote location using the $@$ operator.
When at $p$, $\at{q}{c}$ means

\begin{center}
  \includegraphics{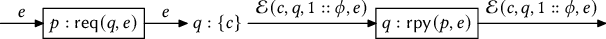}
\end{center}

\begin{figure}

  \[\begin{array}{r@{~=~}l@{\quad}l}
  \evalC{\ast p : t}&\eval{t}{p}{\seq{}}{\mt}\\[2ex]
  \eval{m~q~t}{p}{\phi}{e}&\cnf{m}(p,m,q,t,v_\phi,e)\\
  \eval{\nullify}{p}{\phi}{e}&\mt\\
  \eval{\copyit}{p}{\phi}{e}&e\\
  \eval{\sign}{p}{\phi}{e}&\signp{e}{p}\\
  \eval{\hash}{p}{\phi}{e}&\hashp{e}{p}\\
  \eval{\at{q}{t}}{p}{\phi}{e}&\eval{t}{q}{1\cons\phi}{e}\\
  \eval{t_1\linseqe t_2}{p}{\phi}{e}&\eval{t_2}{p}{2\cons\phi}{
    \eval{t_1}{p}{1\cons\phi}{e}}\\
  \eval{t_1\braseqe{l}{r} t_2}{p}{\phi}{e}
  &\eval{t_1}{p}{1\cons\phi}{\evalF{l}{e}}\sseq
  \eval{t_2}{p}{2\cons\phi}{\evalF{r}{e}}\\
  \eval{t_1\brapare{l}{r} t_2}{p}{\phi}{e}
  &\eval{t_1}{p}{1\cons\phi}{\evalF{l}{e}}\spar
  \eval{t_2}{p}{2\cons\phi}{\evalF{r}{e}}\\[2ex]
  \evalF{d}{e}&\left\{
  \begin{array}{ll}
    \mt&\mbox{if $d=\nil$}\\
    e&\mbox{if $d=\cpy$}
  \end{array}
  \right.
  \end{array}\]

  \caption{Evidence Semantics}\label{fig:evidence semantics}

\end{figure}

Phrases $c_1$ and $c_2$ can be combined using branching.  There are
two ways of combining phrases using branching, sequentially
($\linseq$) and in parallel ($\linpar$). A Copland phrase specifies
whether or not evidence is sent along each branch.  Syntactically,
this is done by writing $\mathbin{d_1 \linseq d_2}$ or $\mathbin{d_1
  \linpar d_2}$ with $d_i \in \{\cpy, \nil\}$ on either side of the
branching operator. If $d_i = \cpy$, $e$ is sent as input evidence
along the corresponding branch.  If $d_i = \nil$, no evidence is sent
along that branch.  We show the pattern for $c_1
\mathbin{\braseqe{\cpy}{\nil}} c_2$ executing at $p$.

\begin{center}
  \includegraphics{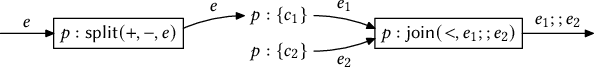}
\end{center}

In the above, $e_i = \eval{c_i}{p}{i::\phi}{d_i(e)}$. Since no
evidence is passed along the branch where $d_i = \nil$, there is no
data flow there and hence no arrow.  In the case of
$c_1\braseqe{\nil}{\nil}c_2$, there is no data flow from any prior
events to the events of $c_1$ or $c_2$.

The pattern for $c_1 \mathbin{\brapare{d_1}{d_2}} c_2$ is identical
except for the label of the last event and the resulting evidence.  In
this case, the last event has label
$p:\cnf{join}(\brapare{}{},e_1\spar e_2)$ and the resulting evidence
is $e_1\spar e_2$.  The sequential and branching operators differ in
terms of their control flow semantics (hence the inclusion of both in
the language): the former ensures the left phrase completes before the
right one begins.  However, the data flow semantics of the two
branching operators are identical.

The following definition introduces notation used to define the data
flow semantics in the appendix.

\begin{defn}[Data Flow Semantics]
  \label{def:data flow semantics}
  The \emph{data flow semantics} of a Copland phrase $\ast p: t$ is
  denoted $\dataC{\ast p : t}$. It relies on an auxiliary function
  $\dataE{t}{p}{\phi}{e}$ which takes a Copland phrase $t$, a place
  $p$, a position $\phi$, and evidence $e$ and produces a data flow
  graph. The detailed inductive definition is shown in
  Figure~\ref{fig:data flow semantics} in the appendix.
\end{defn}

\mypar{Evidence and chain of custody.}  The structure of evidence (see
Fig.~\ref{fig:evidence syntax}) contains ``metadata'' that records
information about how the evidence was assembled from various
pieces. For example, the fact that two pieces of evidence $e_1$ and
$e_2$ were collected in sequence is notated as $e_1\sseq e_2$. If they
were generated concurrently, the resulting evidence would be $e_1\spar
e_2$. Similarly, the structure of a measurement $\cnf{m}(p, m, q, t,
v_\phi, e)$ consists mostly of metadata specifying who the measurer
and the target are. All this metadata can be pre-computed based on the
structure of the Copland phrase being executed. These pre-computed
portions act as the documentation of the chain of custody of the
evidence. The fact that it can be pre-computed means that any attempt
to alter the structure of evidence will be detected by the appraiser,
who is expecting it to look a certain way.

The only part of evidence that cannot be pre-computed is the value
taken by the variable $v_\phi$ in a measurement. It is the variable
$v_\phi$ that holds the actual result of measurement that should
reflect the current state of the target component. Any incriminating
evidence will be found in these $v_\phi$ variables, hence those are
the parts of the evidence structure that an adversary may try to
tamper with.

\mypar{Notation and Terminology.}
Every event occurs at the place found to the left of the colon in its
label.  Since communication events (i.e. request or reply events)
represent the coordinated transmission and reception of data between
places, we identify a sending place at which these events occur as
well as a ``receiving'' place that receives the evidence transmitted
by the sending place.  Although non-communication events do not have a
receiving place, from a proof standpoint it is convenient to simply
define their receiving places to be their sending places.  The
place(s) associated with an event are obtained by place projection
functions.

\begin{defn}[Place Projections]
  \label{def:place projection}
  Let function $\projp$ be the function that extracts the place out of
  an event's label, the place before the colon.

  For an event $v$, define
  \begin{align*}
    \dprojp(v) =
    \begin{cases}
      q & \mbox{$\ell(v)=p:\cnc{req}(q,e)$ or $\ell(v)=p:\cnc{rpy}(q,e)$,}\\
      \projp(v) & \mbox{otherwise.}
    \end{cases}
  \end{align*}
  For a communication event $v$, we call $\projp(v)$ the \emph{sending
    place} and $\dprojp(v)$ the \emph{receiving place} of $v$.
\end{defn}

As a small technical note, Copland syntax does not require that all
requests go to a different place. That is, it is possible for the
sending place and the receiving place of a communication event to be
the same.

\begin{defn}[Cross-place communication]
  Event $v$ is a cross-place communication event iff $\projp(v) \ne
  \dprojp(v)$.
\end{defn}

Finally, we often have the need to refer to the evidence emitted at an
event. The syntax and semantics interact to ensure that this evidence
is always the last argument in the label of an event. We can thus
easily project out this information.

\begin{defn}[Evidence Projection]
  \label{def:evidence projection}
  Let function $\proje$ be the function that projects the last
  argument out of an event's label. This is always the outgoing
  evidence from that event.
\end{defn}

We end this section with a small lemma, a simple sanity check to
ensure the consistency of the data flow and evidence semantics.  The
lemma says that the evidence of the output event of the data flow
semantics is the same as the evidence computed by the evidence
semantics on the same inputs.

\begin{lem}[Flow Graph Top]
  \label{lem:flow graph top}
  If $\dataE{t}{p}{\phi}{e}=D$ then
  $\proje(\top(D))=\eval{t}{p}{\phi}{e}$.
\end{lem}


\section{Motivation}\label{sec:motivation}

In this section, we motivate the study of chain of custody for layered
attestations using a series of example attestation scenarios, modeled
as Copland phrases in the syntax of Figure~\ref{fig:copland syntax}.

Consider a setting in which an appraiser \cnc{app} must assess the
runtime corruption state of a system \cnc{sys} running on a target
device.  The appraiser can request that virus-checking software
\cnc{vc} running in userspace \cnc{us} on the target take a runtime
measurement of \cnc{sys}, which also runs in userspace.  The appraiser
can also request that a trusted virus-checker measurer \cnc{vcm}
running in kernelspace \cnc{ks} take a runtime measurement of
\cnc{vc}.

One way of structuring this attestation would see the \cnc{vcm} first
measure the \cnc{vc} and then hand its evidence off to the latter for
incorporation into the measurement of \cnc{sys}.  The Copland phrase
in Example~\ref{exa:one} represents this structure.

\begin{xmpl}\label{exa:one}
  \[\ast\cnc{app}:\at{\cnc{ks}}{\mbox{\cnc{vcm} \cnc{us} \cnc{vc}}
      \linseqe \at{\cnc{us}}{\mbox{\cnc{vc} \cnc{us} \cnc{sys}}}}\]
\end{xmpl}

Suppose an adversary has corrupted the runtime state of \cnc{vc}
sometime prior to the attestation.  The measurement evidence produced
by the \cnc{vcm} would reveal this corruption to the appraiser.
Unfortunately, in this case the only copy of this evidence the
appraiser receives is that which passes through the corrupt \cnc{vc}.
The adversary in control of \cnc{vc} can direct it to alter the
legitimate, corruption-revealing measurement evidence so that the
appraiser concludes all is well.

The diagram below illustrates the Copland data flow semantics of this
attestation and depicts how evidence tampering at \cnc{vc} allows the
corruption to go undetected.  Corruption-revealing evidence is shown
in bold red, evidence that passes appraisal is in green.

\begin{figure}[h]
  \includegraphics[scale=.71]{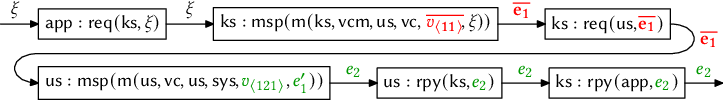}
  \centering
\end{figure}

Since \cnc{vc} receives a copy of the evidence collected by \cnc{vcm},
the \cnc{vc} measurement represents a \emph{tampering opportunity} for
this evidence.  To take advantage of the tampering opportunity, an
adversary must ensure that \cnc{vc} is corrupt when the evidence is
received and that the evidence is not integrity protected in
a way that prohibits tampering.  Integrity protection is discussed
further throughout the rest of the paper. This scenario exemplifies a
conflict of interest: a component whose corruption state is at issue
is nevertheless made sole custodian of the evidence establishing that
state.  Indeed, in this scenario, the adversary need only tamper at
\cnc{vc} to completely avoid detection.

We might attempt to solve this problem by simply providing the
appraiser with an additional copy of the \cnc{vcm} evidence via a
third-party.  This is the approach taken in the attestation of
Example~\ref{exa:two}.  Here, the appraiser also reviews the target
device's asset inventory \cnc{ai}, which is measured by a userspace
asset inventory manager \cnc{aim}.  As indicated by the
$\brapare{+}{+}$ operator, the \cnc{aim} and \cnc{vc} measurements
occur in parallel and both receive the \cnc{vcm} measurement evidence
as input.

\begin{xmpl}\label{exa:two}
	\[\ast\cnc{app}:\at{\cnc{ks}}{\mbox{\cnc{vcm} \cnc{us} \cnc{vc}}
       \linseqe \at{\cnc{us}}{\mbox{\cnc{aim} \cnc{us} \cnc{ai}}
	\brapare{+}{+} \mbox{\cnc{vc} \cnc{us} \cnc{sys}}}}\]
\end{xmpl}

The following diagram depicts a possible execution of this attestation
in which \cnc{vcm} once again collects evidence revealing corruption
at \cnc{vc}.  Separate copies of this evidence $e_1$ reach \cnc{vc}
and \cnc{aim}, making tampering opportunities of both measurements.
Since \cnc{vc} is corrupt, it is in a position to tamper with the copy
of $e_1$ it receives.  However, in order to tamper with the copy that
reaches \cnc{aim}, the adversary must corrupt another component,
namely \cnc{aim} itself or one of the components that receives the
evidence generated by \cnc{aim}.  If the adversary is unable to do so,
this second copy of $e_1$ will ultimately reach the appraiser,
revealing the corruption of \cnc{vc}.

\begin{figure}[h]
  \includegraphics[scale=.71]{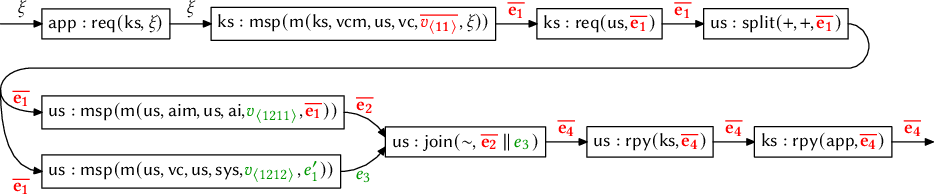}
  \centering
\end{figure}

Thus, the effect of providing several copies of evidence to the
appraiser along separate \emph{data flow paths} is to force an
adversary to work harder in order to avoid detection.  They might be
required to tamper with multiple copies of evidence, that is, take
advantage of multiple tampering opportunities for the evidence,
requiring in turn multiple runtime corruptions in distinct components.
As some of these corruptions may be more feasible than others, the
adversary will be obliged to consider different patterns of corruption
and tampering, which we refer to as alternative \emph{tampering
strategies}.  Of course, an adversary will be interested in
\emph{minimal} tampering strategies, those involving the fewest
corruptions.

Unfortunately, providing redundant copies of evidence is not a
panacea: \cnc{aim}, for instance, may be easy to corrupt, giving
the adversary an easy out.  A more robust approach to making tampering
more difficult is to apply integrity protections to evidence.  The
Copland phrase in Example~\ref{exa:three} is another variation of the
phrase in Example~\ref{exa:one}.  Here, the measurement evidence
generated by \cnc{vcm} is digitally signed with the kernelspace
signing key before it is forwarded to \cnc{vc}.

\begin{xmpl}\label{exa:three}
	\[\ast\cnc{app}:\at{\cnc{ks}}{\mbox{\cnc{vcm} \cnc{us} \cnc{vc}}
	\linseqe \sign \linseqe \at{\cnc{us}}{\mbox{\cnc{vc} \cnc{us} \cnc{sys}} \linseqe \sign}}\]
\end{xmpl}

The userspace component \cnc{vc} cannot access the kernelspace signing
component.  This means an adversary active at \cnc{vc} will be unable
to obtain a new kernelspace signature on tampered evidence.  Any
tampering in spite of this fact will cause a signature verification
failure at the appraiser, alerting it to the adversary's activities
and hence defeating the purpose of tampering in the first place.  In
short, the addition of the kernelspace signature eliminates the
tampering opportunity at \cnc{vc} with the \cnc{vcm} evidence.  The
tampering opportunities in the previous examples crucially depended on
the \cnc{vcm} evidence lacking integrity protections.

As illustrated in the following diagram, evidence of corruption at
\cnc{vc} now makes it all the way to the appraiser regardless of
userspace corruptions.

\begin{figure}[h]
  \includegraphics[scale=.71]{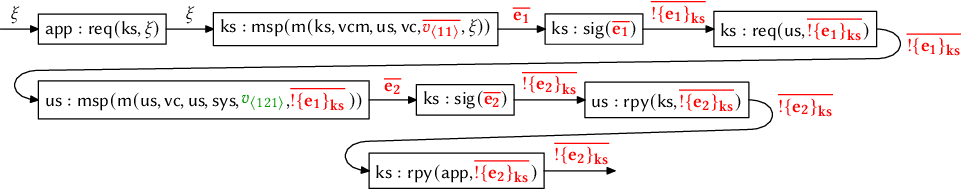}
  \centering
\end{figure}

Thus, the addition of digital signatures at key places in the data
flow can limit---maximally, as we show---the number of tampering
opportunities and strategies available to an adversary.


\section{Paths and Tampering}\label{sec:tampering}
Section~\ref{sec:motivation} illustrated the central role of data flow
analysis in identifying tampering opportunities and strategies. In
this section, we develop the relevant theory to make these analyses
precise.

For the remainder of this section, we fix a data flow graph
$D=\dataE{t}{p}{\phi}{e}$ for some $t$, $p$, and~$e$. Events $v$ and
$v'$ range over $D^V$.

\subsection{Defining Tampering Opportunities}

A component can only tamper with evidence it receives.  Thus, the
first step in identifying tampering opportunities for a piece of
evidence is to identify which components receive a copy of the
evidence as the attestation executes.  Using the data flow semantics,
we can approach this task systematically by considering the data flow
paths beginning at the event that generates the evidence of interest.

\begin{defn}[Path]\label{def:path}
  A \emph{path} of a data flow graph $D = (V, v^i, v^o, \to, \ell)$ is
  a sequence of events $\pi = \seq{v_1,\ldots,v_n}$ such that, for
  every $i<n$, $v_i\to v_{i+1}$. To append an event to the end of a
  path we write $\seq{v_1,\ldots,v_n}\append v_{n+1}$ for
  $\seq{v_1,\ldots,v_{n+1}}$. Concatenation is written as
  $\seq{v_1,\ldots,v_n}\app\seq{w_1,\ldots,w_k} =
  \seq{v_1,\ldots,v_n,w_1,\ldots,w_k}$.  We use $\Pi(v,v')$ to denote
  the set of all paths from $v$ to $v'$.
\end{defn}

Data flow paths represent the trajectories of evidence through the
attestation.  Since the evidence emitted at a non-split data flow
event always includes the evidence received as input there, every
subsequent event along a path beginning with $v$ receives the evidence
emitted at $v$.  This evidence is received at $v'$ if and only if
there is a data flow path starting at $v$ and ending at $v'$.

As discussed in Section~\ref{sec:motivation}, the primary defense
against tampering is to integrity-protect evidence by signing it.  We
assume that each place has a distinct evidence-signing function that
all and only components residing at that place may access.  One
consequence of this assumption is that a signature generated at place
$p$ does not protect against tampering by other components at $p$.
This is because components located at $p$ can always obtain a new $p$
signature on tampered evidence.  Thus, it is important to keep track
of which places have signed evidence as it flows along paths.  An
important case of this is when all signatures (if any) were generated
at a single place.

\begin{defn}[Place-signing path]
  Let $p$ be a place. A path $\pi$ is a \emph{$p$-signing path} iff
  for all $1\leq i\leq \length{\pi}$,
  $\ell(\pi(i)) = q : \cnf{sig}(e)$ implies $q = p$.
\end{defn}

With these definitions, we are ready to formalize the intuitions
developed in Section~\ref{sec:motivation} about an adversary's ability
to tamper with evidence in such a way that an appraiser would not
detect it.

\begin{defn}[Tamper Opportunity]\label{def:tamper-permitting path}
  A path $\pi=\seq{v,\ldots,v'}$ \emph{permits tampering} at $v'$ with
  the evidence from $v$ iff
  \begin{enumerate}
  \item $v$ is a measurement event,
  \item $v\neq v'$ (i.e. $\length{\pi} > 1$), and
  \item $\pi$ is $\projp(v')$-signing or $\dprojp(v')$-signing.
  \end{enumerate}
  The set of tamper-permitting paths from $v$ to $v'$ is denoted
  $\tmprs{v}{v'}$. We say $v'$ is a \emph{tamper opportunity} for $v$
  iff $\tmprs{v}{v'} \ne \varnothing$.
\end{defn}

Condition 1 of Definition~\ref{def:tamper-permitting path} restricts
adversarial tampering to measurement evidence, that is, evidence
created at measurement events.  This focuses our analyses on the core
objective of a tampering adversary, which is to change measurement
results that reveal corruption into ones that do not.  Measurement
events are the only generators of these key values, with the other
data flow events simply passing them along, bundling them into larger
structures, or applying a cryptographic function to them.  By homing
in on the evidence generation sites, we ensure that we have the
complete history of a measurement in mind when deciding its tampering
opportunities.  Condition 1 also serves the dual purpose of excluding
tampering with other forms of evidence that we consider will always be
detected by an appraiser.  For instance, it excludes tampering with a
hash or signature directly or altering the structure of evidence so
that it does not match the appraiser's expectations.

Condition 2 disallows tampering at the event that generates the
evidence of tampering interest.  Such ``self-tampering'' is much more
akin to evidence forgery, in which a corrupt measurer simply lies
about the state of the target and forges a result that will pass
appraisal.  Prior work has considered this kind of evidence forgery in
the layered attestation setting.~\cite{Rowe16a, RoweRK2021}.

Finally, Condition 3 codifies our assumptions about the integrity
protection provided by signatures.  By inspecting the signature events
along a data flow path that delivers the evidence produced at $v$ to
$v'$, we can determine which places sign that copy of the evidence
before it is received.  If $v'$ occurs at $p$ but a signature event at
a different place $q$ is found along the path, the $q$-signature
prevents $v'$ from tampering.  However, $p$-signatures along the path
provide no defense against tampering at $v'$, and if all such
signatures are $p$-signatures, the evidence produced at $v$ lacks any
integrity protections with respect to $v'$.  The component executing
$v'$ can use its access to the signing component at $p$ to obtain new
signatures on tampered evidence.

Definition~\ref{def:tamper-permitting path} allows us to more formally
state the Tamper Opportunities Problem as follows.

\begin{tops}
  Given a Copland phrase $t$, a place $p$, and a measurement event $v$
  in $\dataC{\ast p : t}$, compute
  \(\opps(v) = \{ v' \mid \tmprs{v}{v'} \ne \varnothing \}.\)
\end{tops}

\subsection{Defining Tamper Strategies}

Definition~\ref{def:tamper-permitting path} formalizes our intuitions
surrounding an adversary's ability to tamper with evidence at a single
event, that is, with a particular copy of evidence at a single
component.  However, the lesson of Example~\ref{exa:two} is that
tampering at a single event is sometimes insufficient to avoid
detection.  Rather, \emph{all} copies of a particular piece of
incriminating evidence that reach the appraiser must be tampered with.

As Lemma~\ref{lem:flow graph top} shows, the evidence that results
from executing a Copland phrase is that emitted by the output event of
the phrase's data flow semantics.  This means the appraiser receives
the evidence emitted at any event $v$ that has a data flow path ending
in the output event, i.e., for which $\Pi(v,\top(D))\neq\varnothing$.
Thus, a successful strategy for an adversary to tamper with evidence
created at $v$ consists of a set of events $v'_i$ such that each path
$\seq{v,\ldots,v'_i}$ permits tampering and every path in
$\Pi(v,\top(D))$ encounters one of the $v'_i$.

\begin{defn}[Tamper Strategy]\label{def:tamper strategy}
  A set of events $S$ is a \emph{tamper strategy for} $v\in D^V$ iff
  for each $\pi\in\Pi(v,\top(D))$, there is a $v'\in S$ such that
  $\pi$ contains $v'$ and the subpath $\seq{v,\ldots,v'}$ of $\pi$
  permits tampering.
\end{defn}

Of course, Definition~\ref{def:tamper strategy} leaves room for wildly
inefficient tamper strategies.  For example, if any tamper strategy
for $v$ exists, then the set of all events is a tamper strategy for
$v$.  ``Big'' strategies tell us little about what events, if any,
\emph{must} be part of a tamper strategy for a given event.  They are
also inherently of less interest to an adversary, who will seek to
minimize the number of additional corruptions needed to avoid
detection.  A more natural notion is therefore that of a minimal
tamper strategy in which removing any of the events of $S$ results in
a set that is no longer a tamper strategy.

\begin{defn}[Minimal Tamper Strategy]\label{def:minimal strategy}
  $S$ is a \emph{minimal tamper strategy for} $v\in D^V$ iff
  \begin{enumerate}
  \item $S$ is a tamper strategy for $v$ and
  \item if $S'$ is a tamper strategy for $v$ and $S'\subseteq S$, then
    $S' = S$.
  \end{enumerate}
\end{defn}

Observe that if $\Pi(v,\top(D))$ is empty, any set of events vacuously
satisfies Definition~\ref{def:tamper strategy}, and hence the empty
set is the unique minimal tamper strategy for $v$.  This is reasonable
because the evidence created at $v$ never reaches the appraiser in
this case, hence no tampering with this evidence is necessary.  In
general, however, there may be numerous minimal tamper strategies.

It is now clear how to formally state the Tamper Strategies Problem in
terms of Definition~\ref{def:minimal strategy}.

\begin{tsps}
  Given a Copland phrase $t$, a
  place $p$, and a measurement event $v$ in $\dataC{\ast p : t}$, find
  all minimal tamper strategies $S$ for $v$.
\end{tsps}

\subsection{Computing Tamper Opportunities}

We propose the following procedure that takes as input a starting event $v$
and outputs the set $O$ of tamper opportunities for $v$. It computes
all paths that start with $v$ and collects tamper opportunities within
each path.

\begin{labeledbox}{Algorithm to solve TOP}%
  \begin{enumerate}[(1)]
  \item Set $P = \{\seq{v}\}$ and $O = \varnothing$.
  \item Take a path $\pi\append v$ from $P$
    (initially $\pi = \seq{}$).
    \begin{enumerate}[(a)]
    \item If $\pi\append v$ is tamper-permitting, add $v$ to $O$.
    \item Otherwise leave $O$ unchanged.
    \end{enumerate}
  \item For each $\pi'=\pi\append v\append v'$ extending
    $\pi\append v$ by a single
    event, add $\pi'$ to $P$.
  \item If $P$ is non-empty, go to step~2, otherwise, return $O$.
  \end{enumerate}
\end{labeledbox}

It is easy to see that this procedure always terminates and outputs $O
= \opps(v)$.  The procedure incrementally builds all paths starting at
$v$, checking the three conditions of
Definition~\ref{def:tamper-permitting path} at each step until the
paths cannot be extended further and recording all paths that are
tamper-permitting along the way.

Line 2(a) of this algorithm to solve TOP requires verifying the
conditions of Definition~\ref{def:tamper-permitting path}.  Though
doing so directly is straightforward, we instead introduce the
\emph{tamper set} function to simplify the process.  This function
applies to paths $\seq{v_1,\ldots,v_n}$ and computes the set of places
that can tamper with measurement evidence generated at $v_1$ by the
time it leaves $v_n$.

\begin{defn}[Tamper set]\label{def:tamper set}
  The \emph{tamper set} function $\ts{\cdot}$ outputs a set of places
  given a path and is defined as follows. Let $\mathcal{P}$ be the set
  of all places.
  \[
   \begin{array}{r@{{}={}}l}
    \ts{\seq{v}} & \left\{
                             \begin{array}{ll}
                               \mathcal{P} & \mbox{if }
                                             \ell(v)
                                             = p:\cnf{msp}(\dots)\\
                               \varnothing & \mbox{otherwise}
                             \end{array}\right.\\
    \ts{\pi\append v} &  \left\{
                        \begin{array}{ll}
                          \{p\} \cap \ts{\pi} & \mbox{if } \ell(v) =
                                                p:\cnf{sig}(e)\\
                          \ts{\pi} & \mbox{otherwise}
                        \end{array}\right.
   \end{array}
   \]
\end{defn}

A tamper set is either the set of all places, a singleton set, or the
empty set.
There are two observations to made about Definition~\ref{def:tamper
  set}. First, if $\pi$ does not start with a measurement event, then
$\ts{\pi} = \varnothing$.  This is in keeping with our exclusion of
non-measurement evidence from tampering considerations.  Secondly, it
is easy to check that the function is weakly decreasing in the
following sense:
\begin{equation*}\label{eqn:ts decreasing}
  \ts{\pi_1\app\pi_2}\subseteq\ts{\pi_1}
\end{equation*}%

Theorem~\ref{thm:ts-tamper-permitting} shows that the tamper set
function can be used to verify whether or not a path is
tamper-permitting.

\begin{thm}\label{thm:ts-tamper-permitting}
  For any path $\pi = \pi'\append v$ that starts at a measurement
  event, $\pi$ is tamper-permitting iff either $\projp(v)$ or
  $\dprojp(v)$ is in $\ts{\pi'}$.
\end{thm}

\subsection{Computing Tamper Strategies}

Recall that a tamper strategy for $v$ is a set of events at which the
evidence of corruption produced at $v$ can be can be changed so the
appraiser detects nothing wrong.
Observe that if $v$ has a tamper strategy, then it has a minimal one.
This is because sets of events form a well-founded partial order under
inclusion.  We also remark that, since tamper strategies are sets of
tamper opportunities, every tamper strategy for $v$ is a subset of
$\opps(v)$, the set of all tamper opportunities for $v$.  Finally, we
assert that $\opps(v)$ itself is a tamper strategy for $v$.  While not
immediately obvious, this is a simple consequence of
Theorem~\ref{thm:strategy exists} in the next section.

Given these observations, our approach to solving TSP is to start with
$\opps(v)$, which we can compute via our algorithmic solution to TOP,
and incrementally climb down the partial order of sets until we find
the subsets of $\opps(v)$ that are minimal tamper strategies for
$v$. More concretely, we propose the following procedure that takes as
input an event $v$ and outputs the set $M$ of minimal tamper
strategies for $v$.

\begin{labeledbox}{Algorithm to solve TSP}%
  \begin{enumerate}[(1)]
  \item Set $\mathcal{S} = \{\opps(v)\}$ and $M = \varnothing$.
  \item Take a tamper strategy $S$ from $\mathcal{S}$.
  \item For every $v\in S$, compute $S' = S\setminus\{v\}$.
    \begin{enumerate}[(a)]
    \item If $S'$ is a tamper strategy, add $S'$ to $\mathcal{S}$.
    \end{enumerate}
  \item If no strategies were added to $\mathcal{S}$ in the above step,
    add $S$ to $M$.
  \item If $\mathcal{S}$ is not empty, go to step~2, otherwise
    return $M$.
  \end{enumerate}
\end{labeledbox}


\section{Limits to Constraining Tamper Strategies}\label{sec:results}

We are now equipped to explore the ways in which we can limit the
tamper strategies available to an adversary.

The assumption that $p$ signatures do not provide protection against
tampering by other $p$-located components has significant implications
for our ability to limit tamper strategies.  It means, for instance,
that if $\pi$ is a path whose constituent events all occur at the same
place, then a measurement event in $\pi$ has a tamper opportunity at
every subsequent event in $\pi$.  We begin by formalizing this
observation.

\begin{defn}[Local Path]\label{def:local path}
  A path $\pi=\seq{v_1,\ldots,v_n}$ in a data flow graph is
  \emph{local} to place $p$ iff $\projp(v_i)=\dprojp(v_i)=p$ for all
  $1<i<n$ and $\projp(v_j)=p$ or $\dprojp(v_j)=p$ for $j\in\{1,n\}$.
\end{defn}

A local path is one that never leaves a given
place. Definition~\ref{def:local path} allows for a path local to
place $p$ to start or end with a cross-place communication event whose
sending or receiving place is $p$.  As anticipated, we can indeed
prove that local paths always permit tampering.

\begin{lem}[Local Path Tampers]\label{lem:local path tampers}
  Let $v\in D^V$ be a measurement event and suppose
  $\pi=\seq{v,\ldots,v'}$ is a path of length greater than one. If
  $\pi$ is local, then $\pi$ permits tampering.
\end{lem}

Might there be a way to structure a Copland phrase that precludes the
existence of tamper strategies in spite of this fact?  Unfortunately,
the answer is negative.  Every path from a measurement event to the
output of a data flow graph has a local prefix which permits
tampering.

\begin{thm}[Strategy Exists]\label{thm:strategy exists}
  For every measurement event $v\in D^V$, if $v\neq\top(D)$, then
  there is a tamper strategy for $v$.
\end{thm}

Theorem~\ref{thm:strategy exists} excludes the corner case in which
the measurement event in question is the output event.  This
corresponds to a scenario in which the appraiser takes a measurement
for its own consumption at the conclusion of the attestation.  Since
no events succeed this measurement, there are of course no tamper
opportunities for it.  However, such a measurement would be unusual.
Appraisers typically do not measure themselves, and measurements of
components on other systems are usually mediated by requests and
replies, which makes them subject to Theorem~\ref{thm:strategy
  exists}.  Thus, tamper strategies cannot be eliminated for most
measurement events, because local events follow them.  This negative
result puts an upper limit on how much protection against tampering is
possible.

We cannot hope to eliminate tampering along local paths.  Is it
possible to ensure these are the only available avenues for tampering?
In other words, are there data flow graphs for which no non-local
paths permit tampering?  This time we can answer in the
affirmative. The following definition specifies a property that a path
or a graph can have to ensure that tampering is confined to occur only
along local paths.

\begin{defn}\label{def:protected}
  Let $\pi = \seq{v_1,\ldots,v_n}$ be a path. We say $\pi$ is
  \emph{protected}, iff either
  \begin{enumerate}
  \item $v_1$ is not a measurement event, or
  \item for every $1<i\le n$, if $v_i$ is a cross-place communication
    event, then $\ts{\seq{v_1,\ldots,v_i}}\subseteq \{\projp(v_i)\}$.
  \end{enumerate}
  We say a data flow graph is protected iff all paths in the graph are
  protected.
\end{defn}

A protected data flow graph should be one in which adversarial
tampering is maximally limited.  That is, protected graphs should have
the property that all tamper-permitting paths are local.  This is
indeed the case, as established by the following theorem.

\begin{thm}\label{thm:protected minimal strategies}
  If $\pi = \seq{v_1,\ldots,v_n}$ is protected and permits tampering,
  then $\pi$ is local to $\projp(v_1)$.
\end{thm}


\section{Transforming Copland Phrases}\label{sec:adding signatures}

At the end of the previous section, we introduced the concept of a
\emph{protected} data flow graph and showed that these maximally limit
tamper opportunities.  Here, we present an algorithm that transforms
any Copland phrase $\ast p : t$ into another phrase $\ast p : t'$ that
preserves the evidence collected by the original
phrase, but whose data flow graph is protected.  In doing so, we
provide a solution to the EPPP, which we can now formalize in terms of
protected graphs.

\begin{epps}
  Given a Copland phrase $\ast p : t$, produce another phrase $\ast p
  : t'$ whose semantics is a protected graph that preserves the
  evidence collected by $\ast p : t$.
\end{epps}

Informally, evidence preservation means that all events and data
flows in the original semantics are carried over to the transformed
phrase.

Constructing this algorithm involves shifting from the data
flow-centric view of tampering developed in previous sections to an
evidence-centric view.  To illustrate this shift, consider the
evidence term $e = \signp{e_1}{p} \spar \signp{e_2}{q}$, where $e_1$
and $e_2$ are measurements.  By examining the nested structure of
signatures applied to embedded measurements, we can read off right
from $e$ which places can tamper with each measurement.  Here, for
instance, only events occurring at $p$ can tamper with $e_1$ and only
events at $q$ with $e_2$; thus, a tamper opportunity for the
measurements in $e$ is an event that occurs at $p$ or $q$.  The
structure of evidence encodes enough of each measurement's data flow
history to draw such conclusions.

In order to make these observations precise, we define the
\emph{tamper places} function $\etp$.  Given an evidence term $e$,
$\etp(e)$ is the set of places that can tamper with measurement
evidence embedded in $e$.

\begin{defn}\label{def:etp}
  Let $\mathcal{P}$ be the set of all places. Given an evidence value
  $e$, the \emph{tamper places} function returns a possibly empty set
  of places and is defined inductively as follows.
  \[
    \begin{array}{r@{{}={}}l}
      \etp(\mt) & \varnothing\\
      \etp(\cnf{m}(\cnf{msp}(m,q,t), p, \phi, e)) & \mathcal{P}\\
      \etp(\hashp{e}{p}) & \etp(e)\\
      \etp(\signp{e}{p}) & \{p\} \cap \etp(e)\\
      \etp(e_1\sseq e_2) & \etp(e_1)\cup\etp(e_2)\\
      \etp(e_1\spar e_2) & \etp(e_1)\cup\etp(e_2)
    \end{array}
  \]
\end{defn}

The tamper places function codifies our intuitions about the effects
of the various Copland actions on the vulnerability of measurement
evidence to tampering.  An unsigned measurement is tamperable anywhere,
signatures restrict tampering at most to the places that apply them,
and operations that combine evidence have no effect.

We first establish that the evidence-centric view of tampering
represented by $\etp$ has the correct connection to the path-centric
view.  Recall the tamper set function defined in
Section~\ref{sec:tampering}: $\ts{\seq{v_1,\ldots,v_n}}$ computes the
set of places that can tamper with measurement evidence created at
$v_1$ after it has propagated down the given data flow path through
$v_n$.  Theorem~\ref{thm:ts-tamper-permitting} characterizes
tamper-permitting paths in terms of $\ts{\cdot}$.  We now establish
the following relationship between $\etp(\cdot)$ and $\ts{\cdot}$.

\begin{lem}\label{lem:ts-sub-tau}
  If $\pi = \seq{v_1,\ldots,v_n}$ is a path starting with a
  measurement event, then $\ts{\pi} \subseteq \etp(\proje(v_n))$.
\end{lem}

Thus, applying $\etp$ to the evidence emitted at the end of $\pi$
overapproximates $\ts{\pi}$.  This is sufficient for our purposes, as
it allows us to obtain an evidence-centric criterion for deciding
whether a data flow graph is protected.

\begin{lem}\label{lem:protected-of-cpc-sub}
  Given a flow graph $D$,  if every cross-place communication
  event $v$ in $D$ satisfies $\etp(\proje(v)) \subseteq
  \{\projp(v)\}$, then $D$ is protected.
\end{lem}

Lemma~\ref{lem:protected-of-cpc-sub} provides us with a sufficient
condition to aim for in constructing an algorithm to solve the EPPP.
Namely, we can guarantee a Copland phrase's semantics is a protected
graph, and therefore maximally limits tamper opportunities, by
ensuring that $\etp(\proje(v))\subseteq \{\projp(v)\}$ holds for every
cross-place communication event $v$ in its semantics.  This suggests a
procedure that computes $\etp(e)$ for input evidence $e$ received at
each cross-place communication event $v$ and adds a signature
immediately before $v$ only if $\etp(\proje(v)) \not\subseteq
\{\projp(v)\}$.

\begin{defn}[Evidence Protection Program]\label{def:evi-sign-comm}
  The procedure to transform one phrase into another is defined
  inductively as follows.
  \[
    \begin{array}{r@{{}={}}l}
      \overline{\evisc}(\ast p: t) & \ast p: \evisc(t, p, \mt)\\[2ex]
      \evisc(m~q~t, p, e) & m~q~t\\
      \evisc(\copyit, p, e) & \copyit\\
      \evisc(\hash, p, e) & \hash\\
      \evisc(\sign, p, e) & \sign\\
      \evisc(t_1\linseqe t_2, p, e) & \evisc(t_1, p, e)\linseqe
                                      \evisc(t_2, p,
                                      \eval{\evisc(t_1,p,e)}{p}{\seq{}}{e})\\
      \evisc(\at{p}{t}, p, e) & \at{p}{\evisc(t, p, e)}\\
      \evisc(\at{q}{t}, p, e) & \left\{
                                \begin{array}{ll}
                                   \at{q}{\evisc(t, q, e)} & \mbox{if }
                                                                     \etp(e)\subseteq\{p\}\\&
                                  \mbox{and } \etp(e_1)
                                  \subseteq\{q\}\\
                                  \at{q}{\evisc(t, q, e)\linseqe\sign}
                                                       & \mbox{if }
                                                         \etp(e)\subseteq\{p\}\\&
                                  \mbox{and }
                                  \etp(e_1)
                                  \not\subseteq\{q\}\\
                                  \sign\linseqe\at{q}{\evisc(t, q,
                                  \signp{e}{p})} & \mbox{if }
                                                   \etp(e)\not\subseteq\{p\}\\&
                                  \mbox{and }
                                  \etp(e_2)
                                  \subseteq \{q\}\\
                                  \sign\linseqe\at{q}{\evisc(t, q,
                                  \signp{e}{p})\linseqe\sign} & \mbox{if }
                                                                \etp(e)\not\subseteq\{p\}\\&
                                  \mbox{and }
                                  \etp(e_2)
                                  \not\subseteq \{q\}
                                \end{array}\right.\\
      \evisc(t_1\mathbin{l\,o\,r}t_2, p, e) & \evisc(t_1, p, \evalF{l}{e})
                                        \mathbin{l\,o\,r} \evisc(t_2,
                                        p, \evalF{r}{e})
    \end{array}
  \]
  \begin{center}
  where $e_1 = \eval{\evisc(t,q,e)}{q}{\seq{}}{e}$ and
  $e_2 = \eval{\evisc(t,q,\signp{e}{p})}{q}{\seq{}}{\signp{e}{p}}$
  \end{center}
\end{defn}

As an example, the Copland phrase in Example~\ref{exa:three} of
Section~\ref{sec:motivation} is the output of Evidence Protection
Program on Example~\ref{exa:one}.

This Evidence Protection Program was designed with two goals in mind.
The primary goal is that it should provide a general
solution to the EPPP by transforming any
Copland phrase into a maximally tamper-resistant version of itself.  A
naive approach to our strategy of adding signatures before cross-place
communication events would do so blindly, always inserting a signature
whether or not it has any effect on tampering.  While this would work,
we can do much better.  Our second goal for Evidence Protection Program
is that it should minimally perturb the input phrase, adding
only those signatures needed to ensure its semantics is a protected
graph.  Thus, a necessary feature of the program is idempotency, to
guarantee that when Evidence Protection Program is applied to its own
output, the phrase is left unchanged.

Several particular aspects of Evidence Protection Program merit
discussion.  First, $\evisc$ only inserts signature events into the
phrase and always does so using the $\linseqe$ operator.  This
guarantees that the semantics of the original phrase is preserved by
the transformation, one of the requirements of a solution to the EPPP.
Note also that signatures are only added at the beginning or end of
$\at{q}{t}$ phrases and only when $q$ is known to be different from
$p$, the place where the attestation is currently executing.  This
ensures that signing events are only added immediately before
cross-place communication events.  Finally, the side conditions in
Definition~\ref{def:evi-sign-comm} ensure that we do not add more
signatures than necessary to ensure the condition of
Lemma~\ref{lem:protected-of-cpc-sub}.  In each of the four main cases,
if the condition is already satisfied, no signature is added.  This
parsimonious approach allows us to establish subsequently that
Evidence Protection Program is indeed idempotent.

Theorem~\ref{thm:tau-sub-sndp} establishes that the transformations
applied by Evidence Protection Program guarantee the data flow
semantics of the transformed phrases meet the conditions of
Lemma~\ref{lem:protected-of-cpc-sub}.  As a simple corollary of these
two lemmas, we conclude that the data flow semantics of a phrase
output by $\overline{\evisc}$ is protected.

\begin{thm}\label{thm:tau-sub-sndp}
  For any phrase $t$, place $p$, evidence $e$, sequence $\phi$, if $v$
  is a cross-place communication event in
  $\dataE{\evisc(t, p, e)}{p}{\phi}{e}$, then
  $\etp(\proje(v)) \subseteq \{\projp(v)\}$.
\end{thm}

\begin{cor}
  For any phrase $\ast p: t$, $\dataC{\overline{\evisc}(\ast p: t)}$
  is protected.
\end{cor}

Thus, Evidence Protection Program provides an algorithm to solve the
EPPP for any input Copland phrase, which we state formally.

\begin{labeledbox}{Algorithm to solve EPPP}
  ~Given $\ast p : t$ as input, return $\overline{\evisc}(\ast p : t)$.
\end{labeledbox}

Finally, Theorem~\ref{thm:idempotent} establishes that Evidence
Protection Program is idempotent.  This proves that the side
conditions in Definition~\ref{def:evi-sign-comm} work as intended to
ensure that the only signatures added are those necessary to guarantee
the data flow graphs meet the conditions of
Lemma~\ref{lem:protected-of-cpc-sub}.

\begin{thm}\label{thm:idempotent}
  The function $\evisc$ is idempotent. That is, for any $t, p, e$, it
  follows that $\evisc(\evisc(t,p,e),p,e) = \evisc(t,p,e)$. Hence
  also $\overline{\evisc}(\overline{\evisc}(\ast p : t)) =
  \overline{\evisc}(\ast p : t)$.
\end{thm}

This shows that Evidence Protection Program meets both of its design
goals, including resolving the EPPP.


\section{Related Work}\label{sec:related}
The foundation of our the analyses in our work is
Copland~\cite{RamsdellEtAl2019}, a language for specifying layered
attestations.  We provide several formal semantics to facilitate
precise reasoning about attestation matters.  A language-based
approach for specifying attestations provides flexibility to adapt to
many different situations and
architectures~\cite{HelbleEtAl2021}. There is growing recognition that
this approach is beneficial in supporting the full range of use cases
that can benefit from remote attestation, from software defined
networking~\cite{SultanaSY2022}, to smart cities~\cite{MoreauCS2023},
to flight planning for unmanned aerial
vehicles~\cite{PetzJA2021}. Such recognition is even appearing in
standards documents~\cite{RATS-arch}.

While such flexibility is useful, it introduces the challenge of
having to determine the trustworthiness of any given protocol. A
variety of efforts have investigated the trustworthiness of remote
attestation. These range from low-level analyses of a ``late launch''
capability~\cite{DattaFGK2009} to meta-analyses of high-level
principles~\cite{UsmanCABV2023}. The ones most related to our current
work are those that explore the possibility that an adversary could
corrupt a component after it is measured in order to successfully fool
an appraiser into accepting an attestation when it should
not~\cite{KhurshidR2023, AberaEtAl2016, QinEtAl2020, Rowe16a,
  RoweRK2021}. This is a time-of-check-time-of-use (TOCTOU)
attack. The chain-of-custody attacks considered in this paper share
some aspects with TOCTOU attacks. They both fool an appraiser due to
adversarial actions that occur after the time of measurement. However,
a key difference is that in a TOCTOU attack, the target of measurement
is assumed to be uncorrupted at the time of measurement, yielding
measurement evidence that is not incriminating. In a chain-of-custody
attack, the target is assumed already to be corrupt at the time of
measurement. This yields incriminating evidence that is tampered with
by a downstream corrupt component.

Flexible attestation managers such as~\cite{Maat2018}
and~\cite{PetzA2019} incorporate a selection policy that dictates
which attestation protocols a system is willing to run in different
scenarios. Since the target and the appraiser may have differing
priorities, they must negotiate a protocol that serves the needs of
both parties~\cite{FritzA2023}. Methods for analyzing the
trustworthiness of attestations, such as the present work and those
cited in the previous paragraph, can provide valuable input into the
negotiation and selection process.


\section{Conclusion}\label{sec:conclusion}
In this paper, we initiated the study of evidence chain of custody for
layered attestations.  Using the data flow semantics of Copland
phrases, we can describe the complete history of components that had
access to each copy of given measurement evidence on its way to an
appraiser.  This semantics allowed us to formally define tamper
opportunities for measurement evidence and to develop a procedure for
identifying all components that could tamper with given measurement
evidence at runtime if corrupted by an adversary.  As illustrated via
examples, the possibility of branching execution means that tampering
with a single copy of evidence is generally insufficient for an
adversary to prevent incriminating measurement evidence from reaching
the appraiser.  This drove us to define tamper strategies, sets of
tamper opportunities for given evidence that, if leveraged, would
allow an adversary to avoid detection via evidence tampering.  We then
developed a procedure for identifying all minimal tamper strategies
for given evidence.

While the ability to identify tamper strategies is useful, an
attestation designer's chief interest lies in reducing or eliminating
tamper strategies.  For this reason, we also developed an Evidence
Protection Program that transforms Copland phrases into maximally
tamper-resistant iterations of themselves.  The procedure preserves
the semantics of the input phrase while maximally limiting tamper
opportunities and strategies within it by adding digital signatures at
key places in the semantics.  As we show, Evidence Protection Program
is idempotent, suggesting that it minimally perturbs the input Copland
phrase.  This permits Copland phrase designers to write their phrases
agnostic of tamper-preventing signatures and use Evidence Protection
Program to apply integrity protections afterward.  Evidence Protection
Program and the procedure for finding minimal tamper strategies can
also be used in tandem during the initial design process to suggest
alternatives with more desirable maximally-constrained tamper
properties.

In the future, it would be fruitful to integrate tamper analyses into
a larger trust analysis engine similar to the one presented
in~\cite{RoweRK2021}. This would provide a more comprehensive
understanding of the ways an adversary can avoid detection by
combining evidence tampering with evidence forgery.


\bibliography{tmpr}
\bibliographystyle{plain}

\appendix
\section{Copland Data Flow Semantics}\label{apdx:semantics}

The Copland data flow semantics associates a data flow graph to a
Copland phrase. As we saw above, the semantics is recursively defined
by stitching together smaller data flow graphs. We therefore need to
specify functions to do this stitching. We rely on two methods
depending on whether or not data is passed from the output of one
graph as input to the next.

The before copy operation links two data flow graphs, passing the
output of one to the input of the next.

\begin{defn}[Before Copy]\label{def:before copy}\sloppypar
  Let $D_1=(V_1,v^i_1,v^o_1,\to_1,\ell_1)$ and
  $D_2=(V_2,v^i_2,v^o_2,\to_2,\ell_2)$ be two event systems where
  $V_1$ and $V_2$ are disjoint.  The before copy operation
  $\bfrcpy$ is
  \[D_1\bfrcpy D_2=(V_1\cup V_2, v^i_1,v^o_2,
  \to_1\cup\to_2\cup\{v^o_1\to v^i_2\},
  \ell_1\cup\ell_2).\]
\end{defn}

The before nil operation combines two data flow graphs but does not
link them with a data flow edge.  That is, the edge $v^o_1\to v^i_2$
is omitted.

\begin{defn}[Before Nil]\label{def:before nil}\sloppypar
  Let $D_1=(V_1,v^i_1,v^o_1,\to_1,\ell_1)$ and
  $D_2=(V_2,v^i_2,v^o_2,\to_2,\ell_2)$ be two event systems where
  $V_1$ and $V_2$ are disjoint.  The before nil operation
  $\bfrnil$ is
  \[D_1\bfrnil D_2=(V_1\cup V_2, v^i_1,v^o_2,
  \to_1\cup\to_2,
  \ell_1\cup\ell_2).\]
\end{defn}

Using these two operations, we can now formally define a function that
takes a Copland phrase and returns the data flow graph associated with
it.

\begin{figure}
  \[\begin{array}{r@{~=~}l@{\quad}l}
  \dataC{\ast p : t}&\dataE{t}{p}{\seq{}}{\mt}\\[2ex]
  \dataS{l}&(\{v\},v,v,\emptyset,\{v\mapsto l\})\\
  \dataE{s_1~q~s_2}{p}{\phi}{e}
  &\dataS{p:\cnf{msp}(s_1,q,s_2,\phi,\cnf{m}(\cnf{ms}(s_1,q,s_2),p,\phi,e))}\\
  \dataE{\at{q}{t}}{p}{\phi}{e}
  &\begin{array}[t]{@{}l}
     \dataS{p:\cnf{req}(q,e)}\bfrcpy\dataE{t}{q}{1\cons\phi}{e}\\
     \quad{}\bfrcpy\dataS{q:\cnf{rpy}(p,\eval{t}{q}{1\cons\phi}{e})}
   \end{array}\\
  \dataE{\nullify}{p}{\phi}{e}&\dataS{p:\cnf{nul}(\mt)}\\
  \dataE{\copyit}{p}{\phi}{e}&\dataS{p:\cnf{cpy}(e)}\\
  \dataE{\sign}{p}{\phi}{e}&\dataS{p:\cnf{sig}(\signp{e}{p})}\\
  \dataE{\hash}{p}{\phi}{e}&\dataS{p:\cnf{hsh}(\hashp{e}{p})}\\
  \dataE{t_1\linseqe t_2}{p}{\phi}{e}
  &\dataE{t_1}{p}{1\cons\phi}{e}\bfrcpy
  \dataE{t_2}{p}{2\cons\phi}{\eval{t_1}{p}{1\cons\phi}{e}}\\
  \dataE{t_1\mathbin{l\,o\,r}t_2}{p}{\phi}{e}
  &(V_5,v^i_5,v^o_5,\to_5,\ell_5)\\[2ex]
  D_1&\dataS{p:\cnf{split}(l, r, e)}\\
  D_2&D_1\bfr{l}\dataE{t_1}{p}{1\cons\phi}{\evalF{l}{e}}\\
  D_3&D_1\bfr{r}\dataE{t_2}{p}{2\cons\phi}{\evalF{r}{e}}\\
  D_4&\dataS{p:\cnf{join}(o,\eval{t_1\mathbin{l\,o\,r}t_2}{p}{\phi}{e})}\\
  V_5&V_2\cup V_3\cup V_4\\
  v^i_5&\bot(D_1)\\
  v^o_5&\top(D_4)\\
  \to_5&\to_2\cup\to_3
  \cup\,\{v^o_2\to v^i_4,v^o_3\to v^i_4\}\\
  \ell_5&\ell_2\cup \ell_3\cup \ell_4
  \end{array}\]

  \caption{Data Flow Semantics}
  \label{fig:data flow semantics}
\end{figure}

Definition~\ref{def:data flow semantics} is the formal instantiation
of the informal semantics described in the previous section.


\end{document}